\documentclass[letter]{aa} 

\usepackage{graphicx}
\usepackage{txfonts}
\usepackage[colorlinks=true]{hyperref}
\hypersetup{linkcolor=blue, citecolor=blue}
\usepackage[draft]{todonotes}
\usepackage{mathtools}
\usepackage{booktabs}
\usepackage{multirow}

\usepackage{xcolor}
\definecolor{kostas}{rgb}{1.00, 0.00, 0.0}
\definecolor{vaso}{rgb}{0.00,1.00,0.00}

\begin{document}

   \title{Dancing with the stars: \\Stirring up extraordinary turbulence in Galactic center clouds}
    \titlerunning{Cloud turbulence from stars at Galactic Center}

   \author{Konstantinos Tassis \inst{1} \fnmsep \inst{2}\fnmsep\thanks{tassis@physics.uoc.gr}, Vasiliki Pavlidou \inst{1} \fnmsep \inst{2} \fnmsep\thanks{pavlidou@physics.uoc.gr}
          }

   \institute{
        Department of Physics \& Institute of Theoretical and Computational Physics, University of Crete, GR-70013, Heraklion, Greece
        \and
        Institute of Astrophysics, Foundation for Research and Technology-Hellas, Vasilika Vouton, GR-70013 Heraklion, Greece
        }
    \authorrunning{K. Tassis and V. Pavlidou 2022}
    
   \date{Received ; accepted }

 
  \abstract
   {Molecular clouds in the central molecular zone (CMZ) have been observed to feature turbulent line widths that are significantly higher, and scale with cloud size more steeply, than in the rest of the Milky Way. In the same Galactic region, the stellar density is also much higher than in the rest of the Milky Way, and the vertical stellar velocity dispersion is large, meaning that even young stars are likely to cross the entire vertical extent of the CMZ within their lifetimes.}
   {We investigate whether interactions of CMZ molecular clouds with crossing stars can account for the extraordinary properties of observed turbulence in this part of the Galaxy. }
   {We calculated the rate of energy deposition by stars crossing CMZ clouds due to (a) stellar winds and (b) dynamical friction, and compared it to the rate of turbulence decay. We calculated the predicted scaling of turbulence line width with cloud size in each case.}
   {We find that energy deposition by stellar winds of crossing massive stars can account for both the level and the scaling of CMZ cloud turbulence  with cloud size. We also find that the mechanism stops being effective at a Galactocentric distance comparable to the CMZ extent. On the other hand, we find that dynamical friction by crossing stars does not constitute a significant driver of turbulence for CMZ clouds.  }
   {}


   \maketitle
%

\section{Introduction}

The central molecular zone (CMZ) is an extreme environment in the Milky Way. Turbulence in molecular clouds is extraordinarily high, with line widths of 10-100 km/s, and a scaling with effective cloud size steeper than in the rest of the Milky Way ($\sigma_v \propto L^{0.66 \pm 0.18}$, \citealp{KauffmannEtal2017i}). The driver of this excess turbulence is unclear. Cloud--cloud collisions have been suggested as a possible cause, and the shocks associated with such collisions could also explain the shock chemical tracers found in the Galactic center region  \citep{Menten2009, Lis2001}. However, it is not evident that cloud--cloud collisions could also offer a natural explanation for the steep line--width--size scaling of CMZ turbulence, while the shock chemical tracers could also be the result of hot expanding shells originating in Wolf-Rayet stars \citep{Martin-Pintado1999}.

The stellar density in the Galactic center regions is also very high, including a significant population of massive stars \citep{BryantKrabbe, WRDist,KauffmannEtal2017i}. Stars in the CMZ show significant vertical velocity dispersion ($\sim 50$ km/s; e.g., \citealp{KimMorris2001}). Given that an object moving at 1 km/s crosses 1 pc in 1 Myr, these velocities are sufficiently high even for massive stars to cross the entire vertical extent of the CMZ within their lifetimes. 
 
In this paper, we investigate whether or not stars crossing molecular clouds could be responsible for maintaining the high levels of turbulence in the CMZ, and if so, by which mechanism. The interaction between stars and molecular clouds near the Galactic center has been investigated for its effect on the stellar population;  \citet{KimMorris2001} for example, investigated whether the scattering of stars off of giant molecular clouds in the CMZ  can drive the vertical diffusion of newly formed stars in the region. 
Here, we are looking into the reverse process, that is, whether the presence of a high number of stars affects clouds by driving turbulence within them, via (a) direct kinetic energy deposited by crossing stars that feature fast stellar winds or (b) dynamical friction.

\section{Energetics of turbulence in the CMZ}

Turbulence in molecular clouds is supersonic and generates shocks
that dissipate the energy of the motions on a timescale of the order of a cloud-crossing time \citep{Stoneetal1998}. To maintain a certain level of turbulence, a driver is therefore needed that will replenish the lost energy at a comparable rate. 
The decay rate of kinetic  energy in turbulent motions of typical velocity $v_{\rm turb}$ in a cloud of average density $\rho$ and size $L$ is $\dot{E}_{\rm turb} = \frac{1}{2}\rho L^3 v_{\rm turb}^2 / t_{\rm cross}$, where $t_{\rm cross} = L/v_{\rm turb}$ is the turbulence crossing time. This then gives $\dot{E}_{\rm turb} = \frac{1}{2}\rho L^2 v_{\rm turb}^3$ or, by normalizing to values typical for the CMZ, 
\begin{equation}\label{first}
\dot{E}_{\rm turb} \sim  2\times 10^{35} {\rm \, erg/s} \, \left(\frac{L}{1 \rm \, pc}\right)^2 \left(\frac{n_{H_2}}{10^4{\rm \, cm^{-3}}}\right) 
\left(\frac{v_{\rm turb}}{10 \rm \, km/s}\right)^3 \,.
\end{equation}

\section{Energy deposition by stellar winds in the CMZ}

The Galactic center is a region of high stellar density, with a high abundance of 
stars that produce significant winds, of the order of $10^{-3}{\rm \, M_\odot/yr}$ within the central parsec \citep{Coker2001}. 
Clouds in the CMZ will interact with these stars, both young and older. Stars will have significant relative velocities with respect to molecular clouds, and will be crossing these clouds, stirring turbulence in them in two ways: dynamical friction, and cloud--wind interaction. 
In this section, we calculate the latter. 

Between the central parsec and a Galactocentric distance $R \sim 100 - 300$ pc typical of the CMZ , the self-consistent model of \citet{SormaniEtal2021} for the inner Milky Way bulge predicts a decline of the stellar population by a factor of $\sim 10^3 - 10^4$. Assuming the integrated wind mass-loss rate will decline by the same factor, and adopting the conservative estimate of a $\times 10^4$ decline, we expect a typical mass deposition due to stellar winds of $10^{-7}{\rm \, M_\odot\, yr^{-1} pc^{-3}}$. For stars that happen to be crossing clouds in the CMZ, these winds will deposit a kinetic energy of $\sim 5\times 10^{42} {\rm \, erg\,  yr^{-1} pc ^{-3}}$ (using a conservative average energy deposition rate over the lifetimes of large stars of $\sim 10^{49} {\rm \, erg}$ per $M_\odot$ of lost mass\footnote{From a comparison of top and bottom panels of Fig. 1 of \citet{FierlingerEtal2016} it follows that kinetic energy injection rate and mass-loss rate scale with stellar mass in approximately the same way for massive stars ($>20{\rm \, M_\odot}$), implying a wind velocity roughly constant with mass for these stars; this is why in Eq.~(\ref{second}) it is sufficient to scale the kinetic energy injection rate with the mass wind loss rate alone.} from the numerically integrated models of \citealp{FierlingerEtal2016}, who used tabulated mass-loss rates from \citealp{eks} and wind velocities from \citealp{voss}). Normalizing again, as in Eq.~\ref{first}, to a cloud of size $\sim 1 {\rm pc,}$ this implies an energy deposition rate by winds of 
\begin{equation}\label{second}
 \dot{E}_{\rm kin, winds} = 2\times 10^{35}  {\rm erg/s} \left(\frac{\dot{M}_{\rm winds}}{10^{-7} {\rm \, M_\odot\,yr^{-1} pc^{-3}}}\right)  \left(\frac{L}{\rm 1pc}\right)^3 \,, 
\end{equation}
which, very suggestively, lies at exactly the right level to replenish turbulence decaying at the rate of Eq.~(\ref{first}).

The least secure assumption in deriving Eq.~(\ref{second}) is that the number of massive stars producing strong winds scales in the same way as the total stellar mass between the central parsec and the typical CMZ radii. For this reason, we also present an alternative calculation of $\dot{E}_{\rm kin, winds}$ with different, although still large, uncertainties. 

From \citet{FierlingerEtal2016} (their Fig. 1), the (averaged over stellar lifetime) kinetic energy injection rate of a single star as a function of stellar mass can be approximately described ---for stars more massive than $20{\rm \, M_\odot}$--- by $\dot{E}_{\rm kin, 1} (m) \approx (3.5\times 10^{37}{\rm erg/yr} )\, m^4$, where $m$ is the stellar mass in ${\rm M_\odot}$.  From the same figure, stellar lifetimes {\em in the same mass range} (including the post-main sequence wind stage) can be approximated by $T (m) \approx (30 {\rm Myr}) \, m^{-0.45}$.

If stars are produced in the CMZ at a rate of $\dot{M}_*$ (in ${\rm M_\odot/yr}$, assumed constant over the past few tens of millions of years) with a power-law initial mass function\footnote{We normalize the IMF so that $\int_{m_{min}}^{m_{max}} m (dn/dm)dm =\dot{M}_*$. This normalization yields $C = \dot{M}_*(s-2)/(m_{min}^{2-s}-m_{max}^{2-s})$, 
so that $dn/dm$ itself has units of number of stars per ${\rm M_\odot}$ per yr.}  (IMF)  $dn/dm = C m^{-s}$, then the number of (massive) stars with masses between $m$ and $m+dm$ will be $T(m)(dn/dm)dm$ and their kinetic energy injection rate over the entire CMZ will be $\dot{E}_{kin,1}T(m)(dn/dm)dm$. Integrating over all stars with $m>20$, we obtain a total $\dot{E}_{\rm kin, winds}$ from such stars of $\int_{m_{min}}^{m_{max}} \dot{E}_{kin,1}T(m)(dn/dm)dm$.
Adopting standard parameters for a Salpeter IMF ($s=2.35$,  $m_{max}=120$, and $m_{min}=0.15$ so that $\langle m \rangle \sim 0.5$), and a $\dot{M}_* \sim 0.1$ (e.g. \citealp{Henshaw2022} and references therein) we obtain $\dot{E}_{\rm kin, winds} \sim 10^{40} {\rm erg/s}$ for the entire volume of the CMZ to which the star formation estimate refers. Converting this to a per-volume estimate is not trivial. First, the star formation is not uniformly distributed over the CMZ volume, as is clear from \citet{Henshaw2022}, who compare different estimates of $\dot{M}_*$ in the CMZ using different tracers and studying different volumes: as can be seen in their Table 1, there is no strong correlation between the volume of the CMZ studied and the derived value of $\dot{M}_*$, which is to be expected if the star formation, massive stars, and molecular clouds are all spatially correlated and only cover a small fraction of the CMZ volume. If we assume that most of this star formation takes place in the same region of the central CMZ covered by the clouds studied in \citet{KauffmannEtal2017i} (radius of $\sim 100$ pc, height of $\sim \pm 5$ pc from the Galactic plane), then we obtain $\dot{E}_{\rm kin, winds} \sim 4\times 10^{34}\, {\rm erg \, s^{-1} pc^{-3}}$ (corresponding to a population of about ~1500 O and B stars of all evolutionary stages within the CMZ). However, this estimate is very uncertain. It is most sensitive to our assumptions on the IMF and the volume within most of the star formation is confined. For example, if we assume a top-heavy IMF similar to that preferred for the central parsec (e.g. $s=0.85$, $m_{min}=0.7$ as in Model 8 of \citealp{Maness2007}), we obtain $\dot{E}_{\rm kin, winds} \sim 6\times 10^{35}\, {\rm erg \, s^{-1} pc^{-3}}$. Conversely, for every factor-two increase in the Galactrocentric radius of the CMZ zone, where most of the star formation takes place, the estimate of $\dot{E}_{\rm kin, winds}$ is decreased by a factor of four. However, it is encouraging that, even with these large uncertainties, this first-principles estimate is of the same order of magnitude as the simple calculation of Eq.~(\ref{second}).

\section{Scaling of turbulent line widths with cloud size}

If the turbulence driving of Eq.~(\ref{second}) balances the decay of 
Eq.~(\ref{first}), then a specific scaling of turbulent velocities, $v_{\rm turb}$, with cloud or clump size, $L$, is predicted. Indeed, if the cloud density scales with radius as $\rho \propto r^{-k}$, then the average number density within a cloud or clump of typical size $L$ will also scale as $n_{H_2} \propto L^{-k}$. Then $\dot{E}_{\rm kin, winds} \sim \dot{E}_{\rm turb}$ implies $L^{2-k}v_{\rm turb}^3 \propto L^3$ or 
\begin{equation}\label{third}
    v_{\rm turb} \propto L^{(1+k)/3}\,.
\end{equation}
\citet{KaufmannEtal2017ii} find, for the majority of structures in the CMZ, $k\sim 1.3$, so Eq.~(\ref{third}) gives $ v_{\rm turb} \propto L^{0.77}$, in excellent agreement with the observed scaling in the CMZ reported by \citet{ShettyEtal2012} (most-probable scaling through Bayesian inference of $v_{\rm turb} \propto L^{0.78}$ with a $95\%$ confidence interval on the slope from $0.41$ to $1.13$) and by  \citet{KauffmannEtal2017i} ($ v_{\rm turb} \propto L^{0.66\pm 0.18}$). 
\section{Galactocentric distance cutoff of enhanced turbulence due to stellar winds}

In our discussion, we assume that the number and distribution of massive, fast-wind stars scales with the total stellar density between the central parsec and the outer edge of the CMZ. 
Interestingly, young stars are observed throughout the CMZ  \citep{BryantKrabbe}.  \citet{KauffmannEtal2017i} do observe O-stars and/or masers in several of the CMZ clouds that they have investigated, but not in all of them. It is therefore interesting to ask whether or not it is necessary to have at least one high-mass star {currently} present in every cloud observed to be abnormally turbulent, if the excess turbulence is indeed produced by winds from crossing stars.  The answer is no:
the present level of turbulence in a cloud could have been driven by a crossing massive star in the past. Stars near the Galactic center have vertical velocity dispersions that are significantly higher ($\sim \times 5$) than  typical turbulence velocities in clouds, and therefore the star crossing time will be shorter than the turbulence crossing time  (on which turbulence decays)  by the same factor. Clouds with extraordinary turbulence driven by this mechanism could spend most of their lifetimes devoid of high-mass, high-wind stars, which is consistent with the lack of evidence for the presence of massive stars in several CMZ clouds. 

We can further quantify this argument, to predict the Galactocentric distance beyond which this mechanism will no longer be efficient in stirring up excess turbulence in molecular clouds. This will occur once the number of stars with significant winds making a cloud crossing over a turbulence decay time $L/v_{\rm turb}$ drops below 1. 
The number of crossings through a cloud of size $L$ by wind-producing stars of number density $n_\star$ and crossing velocity $v_\star$ in this time is $n_\star v_\star L^2 (L/v_{\rm turb})$. For the mechanism to remain effective, we need this to be at least 1, that is, we demand that $n_\star L^3 \geq  v_{\rm turb}/v_\star$. If the suppression factor of stellar density compared to the central parsec as a function of Galactocentric distance $R$ is $f(R)$, and the central parsec has $\sim 10^2$ stars producing significant fast winds, then this mechanism will stop being effective at a Galactocentric distance $R$ such that $f(R) \geq (L/0.2 {\rm pc})^{-3}(v_{\rm turb}/50 {\rm \, km\,s^{-1}})$. For clouds of $L\sim 3 {\rm pc}$ in size with $v_{\rm turb} \sim 10 {\rm \, km \,s^{-1}}$ \citep{KauffmannEtal2017i}, we require $f(R) \gtrsim 10^{-4}$, which corresponds to a Galactocentric distance of $\sim 250 {\rm pc}$ \citep{SormaniEtal2021}, which is comparable to the extent of the CMZ. 

\section{Energy deposition through dynamical friction}
That the level and scaling of the kinetic energy injection by stellar winds matches the observations of turbulence in CMZ clouds is a non-trivial result. To emphasise this point, we present here a calculation of an alternative source of kinetic energy injection.

As stars cross through molecular clouds, they experience gravitational drag, that is, a loss of momentum and energy due to {gravitational} interaction with the surrounding matter. The phenomenon, known as dynamical friction, was first studied by  \citet{Chandra1943}. 
The rate of change of the (relative to the cloud) velocity vector of a star, $\vec{v}_\star$, due to dynamical friction, is (see \citealp{BinneyTremaineBook}, chapter 8.1)
\begin{equation}
\frac{d\vec{v}_\star}{dt} = - \frac{4\pi (\ln \Lambda) G^2\rho M_\star}{v_\star^3}
\left[{\rm erf} (X) - \frac{2X}{\sqrt{\pi}}\exp[-X^2]\right]\vec{v}_\star
,\end{equation}
where 
$M_\star$ is the mass of the star; $\rho$ is the gas density; $G$ is the gravitational constant; $X = v_\star/(\sqrt{2}\sigma)$, where $\sigma$ is the velocity dispersion of the gas; and $\Lambda = b_{\rm max} v_0^2/GM_\star$, where 
 $b_{\rm max}$ is the maximum distance that needs to be considered (of the order of the size of the cloud, and so we take $b_{\rm max} \approx L/2$) and $v_0 \approx v_\star$ is the initial relative velocity vector of the star.
The rate of energy deposit in the cloud by a single star is 
$\dot{E}_{\star,1} = M_\star \vec{v}_\star\frac{d\vec{v}_\star}{dt}$ and the total energy deposition rate by stars due to dynamical friction is $\dot{E}_{\star} = \dot{E}_{\star,1} n_\star L^3$ where $n_\star$ is the number density of stars. For $L$ between $1$ and $100$ pc, and a typical $v_\star$ of 50 km/s (e.g. \citealp{KimMorris2001}),  $\ln \Lambda$ will vary between 12 and 17; we adopt $\ln \Lambda \sim 15$.  As $v_\star$ is much higher than $\sigma$, $X$ will be large, and so the term in the square brackets will be of order unity.
For Galactocentric distances $100 {\rm \, pc} \lesssim R\lesssim 300 {\rm \, pc}$, \cite{SormaniEtal2021} modeled a local stellar density of $10^2 {\, \rm M_\odot \, pc^{-3}} \gtrsim \rho_\star \gtrsim 10 {\, \rm M_\odot \, pc^{-3}}$. Normalizing  to
a $1 {\rm \, pc}$ cloud, as in Eqs.~(\ref{first}) and (\ref{second}), and adopting an average stellar mass of $0.5 M_\odot$, we find
\begin{equation} \label{fourth}
\dot{E}_{\star} \!\!= \!8\times 10^{29} {\rm erg/s} 
\left(\frac{\rho_\star}{\rm 10^2 M_\odot pc^{-3}}\right)
\left(\frac{n_{H_2}}{10^4{\rm cm^{-3}}}\right)
\left(\frac{L}{\rm 1 \, pc}\right)^3\!\!
\left(\frac{v_\star}{50 \rm \, km/s}\right)^{-1}\!\!,
\end{equation}
 over five orders of magnitude too low to be relevant for maintaining the observed levels of turbulence in CMZ clouds. 
Even if the level of energy deposition rate by dynamical friction were comparable to the turbulence decay rate, the predicted scaling in this case would be incompatible with observations: from comparing Eqs.(\ref{first}) and (\ref{fourth}) we obtain $L^2n_{H_2}v_{\rm turb}^3 \propto
n_{H_2}L^3$, implying $v_{\rm turb} \propto L^{1/3}$, much shallower than the observed size--line-width relation in the CMZ \citep{ShettyEtal2012,KauffmannEtal2017i}. We conclude that dynamical friction of clouds on stars does not contribute appreciably to the observed level of turbulence in the CMZ. 

\section{Conclusions}

We show that fast stellar winds from massive stars crossing CMZ clouds can nicely reproduce the observed properties of turbulence in CMZ clouds, including its magnitude and scaling with cloud size. We also show that this mechanism ceases to be effective at a Galactocentric radius comparable to the  extent of the CMZ. In contrast, we find that dynamical friction is not important as a turbulence driver for CMZ clouds in magnitude, nor does it reproduce the observed scaling between line width and size.

\begin{acknowledgements}
We thank Kallia Petraki for the stimulating discussions that led to the idea presented in this paper, and the referee, Jens Kauffmann, for constructive comments that helped improve this manuscript. This project has received funding from the European Research Council (ERC) under the European Unions Horizon 2020 research and innovation programme under grant agreement No. 771282. V.P. acknowledges support from the Foundation of Research and Technology - Hellas Synergy Grants Program through project MagMASim, jointly implemented by the Institute of Astrophysics and the Institute of Applied and Computational Mathematics and by the Hellenic Foundation for Research and Innovation (H.F.R.I.) under the ``First Call for H.F.R.I. Research Projects to support Faculty members and Researchers and the procurement of high-cost research equipment grant'' (Project 1552 CIRCE).
\end{acknowledgements}
\bibliographystyle{aa}
\bibliography{bibliography}

\begin{thebibliography}{19}
\expandafter\ifx\csname natexlab\endcsname\relax\def\natexlab#1{#1}\fi

\bibitem[{Binney {et~al.}(1987)Binney, Tremaine, \&
  Ostriker}]{BinneyTremaineBook}
Binney, J., Tremaine, S., \& Ostriker, J. 1987, Galactic Dynamics, Princeton
  series in astrophysics (Princeton University Press)

\bibitem[{{Bryant} \& {Krabbe}(2021)}]{BryantKrabbe}
{Bryant}, A. \& {Krabbe}, A. 2021, \nar, 93, 101630

\bibitem[{{Chandrasekhar}(1943)}]{Chandra1943}
{Chandrasekhar}, S. 1943, \apj, 97, 255

\bibitem[{{Coker}(2001)}]{Coker2001}
{Coker}, R.~F. 2001, \aap, 375, L18

\bibitem[{{Ekstr{\"o}m} {et~al.}(2012){Ekstr{\"o}m}, {Georgy}, {Eggenberger},
  {Meynet}, {Mowlavi}, {Wyttenbach}, {Granada}, {Decressin}, {Hirschi},
  {Frischknecht}, {Charbonnel}, \& {Maeder}}]{eks}
{Ekstr{\"o}m}, S., {Georgy}, C., {Eggenberger}, P., {et~al.} 2012, \aap, 537,
  A146

\bibitem[{{Fierlinger} {et~al.}(2016){Fierlinger}, {Burkert}, {Ntormousi},
  {Fierlinger}, {Schartmann}, {Ballone}, {Krause}, \&
  {Diehl}}]{FierlingerEtal2016}
{Fierlinger}, K.~M., {Burkert}, A., {Ntormousi}, E., {et~al.} 2016, \mnras,
  456, 710

\bibitem[{{Henshaw} {et~al.}(2022){Henshaw}, {Barnes}, {Battersby}, {Ginsburg},
  {Sormani}, \& {Walker}}]{Henshaw2022}
{Henshaw}, J.~D., {Barnes}, A.~T., {Battersby}, C., {et~al.} 2022, arXiv
  e-prints, arXiv:2203.11223

\bibitem[{{Kauffmann} {et~al.}(2017{\natexlab{a}}){Kauffmann}, {Pillai},
  {Zhang}, {Menten}, {Goldsmith}, {Lu}, \& {Guzm{\'a}n}}]{KauffmannEtal2017i}
{Kauffmann}, J., {Pillai}, T., {Zhang}, Q., {et~al.} 2017{\natexlab{a}}, \aap,
  603, A89

\bibitem[{{Kauffmann} {et~al.}(2017{\natexlab{b}}){Kauffmann}, {Pillai},
  {Zhang}, {Menten}, {Goldsmith}, {Lu}, {Guzm{\'a}n}, \&
  {Schmiedeke}}]{KaufmannEtal2017ii}
{Kauffmann}, J., {Pillai}, T., {Zhang}, Q., {et~al.} 2017{\natexlab{b}}, \aap,
  603, A90

\bibitem[{{Kim} \& {Morris}(2001)}]{KimMorris2001}
{Kim}, S.~S. \& {Morris}, M. 2001, \apj, 554, 1059

\bibitem[{{Lis} {et~al.}(2001){Lis}, {Serabyn}, {Zylka}, \& {Li}}]{Lis2001}
{Lis}, D.~C., {Serabyn}, E., {Zylka}, R., \& {Li}, Y. 2001, \apj, 550, 761

\bibitem[{{Maness} {et~al.}(2007){Maness}, {Martins}, {Trippe}, {Genzel},
  {Graham}, {Sheehy}, {Salaris}, {Gillessen}, {Alexander}, {Paumard}, {Ott},
  {Abuter}, \& {Eisenhauer}}]{Maness2007}
{Maness}, H., {Martins}, F., {Trippe}, S., {et~al.} 2007, \apj, 669, 1024

\bibitem[{{Mart{\'\i}n-Pintado} {et~al.}(1999){Mart{\'\i}n-Pintado}, {Gaume},
  {Rodr{\'\i}guez-Fern{\'a}ndez}, {de Vicente}, \&
  {Wilson}}]{Martin-Pintado1999}
{Mart{\'\i}n-Pintado}, J., {Gaume}, R.~A., {Rodr{\'\i}guez-Fern{\'a}ndez}, N.,
  {de Vicente}, P., \& {Wilson}, T.~L. 1999, \apj, 519, 667

\bibitem[{{Menten} {et~al.}(2009){Menten}, {Wilson}, {Leurini}, \&
  {Schilke}}]{Menten2009}
{Menten}, K.~M., {Wilson}, R.~W., {Leurini}, S., \& {Schilke}, P. 2009, \apj,
  692, 47

\bibitem[{{Rosslowe} \& {Crowther}(2015)}]{WRDist}
{Rosslowe}, C.~K. \& {Crowther}, P.~A. 2015, \mnras, 447, 2322

\bibitem[{{Shetty} {et~al.}(2012){Shetty}, {Beaumont}, {Burton}, {Kelly}, \&
  {Klessen}}]{ShettyEtal2012}
{Shetty}, R., {Beaumont}, C.~N., {Burton}, M.~G., {Kelly}, B.~C., \& {Klessen},
  R.~S. 2012, \mnras, 425, 720

\bibitem[{{Sormani} {et~al.}(2021){Sormani}, {Sanders}, {Fritz}, {Smith},
  {Gerhard}, {Schoedel}, {Magorrian}, {Neumayer}, {Nogueras-Lara},
  {Feldmeier-Krause}, {Mastrobuono-Battisti}, {Schultheis}, {Shahzamanian},
  {Vasiliev}, {Klessen}, {Lucas}, \& {Minniti}}]{SormaniEtal2021}
{Sormani}, M.~C., {Sanders}, J.~L., {Fritz}, T.~K., {et~al.} 2021, arXiv
  e-prints, arXiv:2111.12713

\bibitem[{{Stone} {et~al.}(1998){Stone}, {Ostriker}, \&
  {Gammie}}]{Stoneetal1998}
{Stone}, J.~M., {Ostriker}, E.~C., \& {Gammie}, C.~F. 1998, \apjl, 508, L99

\bibitem[{{Voss} {et~al.}(2009){Voss}, {Diehl}, {Hartmann}, {Cervi{\~n}o},
  {Vink}, {Meynet}, {Limongi}, \& {Chieffi}}]{voss}
{Voss}, R., {Diehl}, R., {Hartmann}, D.~H., {et~al.} 2009, \aap, 504, 531

\end{thebibliography}
\end{document}